\documentclass[pdflatex,sn-mathphys-num]{sn-jnl}% Math and Physical Sciences Numbered Reference Style
\usepackage{graphicx}%
\usepackage{multirow}%
\usepackage{amsmath,amssymb,amsfonts}%
\usepackage{amsthm}%
\usepackage{mathrsfs}%
\usepackage[title]{appendix}%
\usepackage{xcolor}%
\usepackage{textcomp}%
\usepackage{manyfoot}%
\usepackage{booktabs}%
\usepackage{algorithm}%
\usepackage{algorithmicx}%
\usepackage{algpseudocode}%
\usepackage{listings}%
\usepackage[normalem]{ulem}%
\theoremstyle{thmstyleone}%
\theoremstyle{thmstyletwo}%

\theoremstyle{thmstylethree}%

\begin{document}

\title[Article Title]{Probing lattice fluctuations using solid-state high-harmonic spectroscopy}

%%=============================================================%%
%% GivenName	-> \fnm{Joergen W.}
%% Particle	-> \spfx{van der} -> surname prefix
%% FamilyName	-> \sur{Ploeg}
%% Suffix	-> \sfx{IV}
%% \author*[1,2]{\fnm{Joergen W.} \spfx{van der} \sur{Ploeg} 
%%  \sfx{IV}}\email{iauthor@gmail.com}
%%=============================================================%%

\author[1]{\fnm{Lance} \sur{Hatch}}
\equalcont{These authors contributed equally to this work.}
\author[2]{\fnm{Navdeep} \sur{Rana}}
\equalcont{These authors contributed equally to this work.}
\author[3]{\fnm{Shoushou} \sur{He}}
\author[3]{\fnm{Jessica} \sur{Yu}}
\author[4]{\fnm{Boyang} \sur{Zhao}}
\author[4]{\fnm{Yu} \sur{Zhang}}
\author[4]{\fnm{Haidan} \sur{Wen}}
\author[3]{\fnm{Xavier} \sur{Roy}}
\author[5]{\fnm{Lun} \sur{Yue}}
\author[2]{\fnm{Mette} \sur{Gaarde}}
\author[1]{\fnm{Hanzhe} \sur{Liu}}

\affil[1]{\orgdiv{Department of Chemistry}\orgname{, Purdue University}\orgaddress{\city{, West Lafayette}, \postcode{47907}, \state{Indiana}, \country{USA}}}
\affil[2]{\orgdiv{Department of Physics and Astronomy}, \orgname{Louisiana State University}, \orgaddress{\city{Baton Rouge}, \postcode{70803}, \state{Louisiana}, \country{USA}}}
\affil[3]{\orgdiv{Department of Chemistry }, \orgname{Columbia University}, \orgaddress{\city{New York}, \postcode{10027}, \state{New York}, \country{USA}}}
\affil[4]{\orgdiv{Materials Science Division }, \orgname{Argonne National Laboratory}, \orgaddress{\city{Lemont}, \postcode{60439}, \state{Illinois}, \country{USA}}}
\affil[5]{\orgdiv{Department of Physics, Applied Physics and Astronomy}, \orgname{Binghamton University}, \orgaddress{\city{Binghamton}, \postcode{13902}, \state{New York}, \country{USA}}}

%%==================================%%
%% Sample for unstructured abstract %%
%%==================================%%

\abstract{Solid-state high-harmonic spectroscopy allows the study of strongly driven ultrafast electron dynamics.
% is this true? I thought HHG has been demonstrated in amorphous systems as well. You are correct, but I think it's OK to focus on periodic systems here. The sentence as written is correct.
Microscopically, high harmonics are generated by strong-laser-field acceleration of electron-hole pairs through the lattice. 
At finite temperatures, atomic-scale structural fluctuations 
%\sout{that change the local lattice structure }
are ubiquitous and are expected to influence the electron-hole trajectories. Yet, the effect of  thermal lattice fluctuations on solid-state high-harmonic generation (HHG) has not been quantified. 
Here, we demonstrate a profound sensitivity of HHG to thermal lattice fluctuations, by characterizing the temperature dependence of HHG in Re$_6$Se$_8$Cl$_2$, a superatomic semiconductor. 
As the sample temperature is decreased, the high-harmonic yield exhibits a slow increase, followed by an abrupt increase below 50 K, consistent with the temperature at which lattice vibrations are strongly suppressed. 
%Our ab-initio calculations show that at higher temperatures, thermal lattice fluctuations of the lattice suppress the harmonic yield via electronic decoherence between electron-hole pairs in different local crystal environments.  
Our calculations show that thermal lattice fluctuations both weaken the harmonic response from individual distorted configurations and induce phase dispersion across the ensemble, leading to a pronounced suppression of the coherently emitted harmonics.
We show that this effect can be interpreted in terms of an effective electronic dephasing time that varies with temperature. Our results are relevant to dephasing in broad strong-field phenomena, including lightwave electronics and Floquet engineering. The wide tunability of superatomic crystals further enables materials-controlled strong-field physics.
}


\maketitle

\section{Introduction}\label{sec1}
High-harmonic generation (HHG) is an extreme nonlinear optical phenomenon in which an intense laser field drives a nonlinear medium to emit coherent radiation at integer multiples of the fundamental frequency~\cite{ferray1988multiple}. 
In the gas phase, this process provides the foundation for attosecond pulse generation and represents one of the fastest ways to coherently control electron motion~\cite{krausz2009attosecond}. 
Its extension to condensed-phase systems has transformed HHG from a source of ultrashort light pulses into a probe of driven quantum matter, granting access to ultrafast, strongly-driven electron-hole dynamics in the crystalline solids~\cite{ghimire2011observation}. 

%Since its initial discovery in noble gases, HHG has been extended to molecules, liquids, and various solids,  
%In noble gases, high-harmonic generation is described by the three-step model. First, a strong, long-wavelength laser field tunnel ionizes a valence electron into the continuum. Next, the laser field accelerates the electron away from its parent ion. Finally, when the sign of the driving field flips, the laser drives the electron back to recollide with the ion. As a result of this recollision, a coherent extreme ultraviolet (XUV) pulse is emitted. The emitted photon energy is proportional to the ionization potential of the atom plus the electron’s kinetic energy gained from the laser acceleration. The emission from successive laser driving cycles interferes, giving rise to a series of odd high harmonics. This model successfully accounts for key features of gas-phase HHG, including the high-energy cutoff and its scaling with the driving field wavelength.
% which has opened new opportunities for probing strongly driven electron-hole dynamics in the condensed phase~\cite{ghimire2011observation}. 

In solids, the HHG process can be understood within a three-step model similar to the well-known gas-phase description~\cite{corkum1993plasma, kulander1993dynamics, schafer1993above}.
In this model, electron-hole pairs that are created through tunneling are subsequently accelerated in the band structure, and their recombination as well as their intraband acceleration give rise to high-harmonic emission~\cite{Vampa2014Analysis, Mengxi2015Bloch, Schubert2014Sub, Ghimire2019Solids, luu2018measurement, luu2015extreme}. 
%Additionally, other HHG mechanisms, such as nonlinear intraband motion and Berry curvature effects, have been discovered in various systems. 
Using this understanding and the formal description that underlies the nonlinear generation process, HHG  has been used to reconstruct band structures, measure Berry curvature, probe topological phase transitions, and image real-space charge distributions in solids~\cite{vampa2015all, luu2018measurement, bauer2018high, silva2019topological, bharti2022high, narolansky2024berry, lakhotia2020laser, heide2022topological}. 
While these studies have successfully revealed the multifaceted aspects of solid-state HHG, their descriptions are largely based on single-electron or single-hole dynamics, with many-body effects being ignored.

Compared with gas-phase HHG, a defining feature of solid-state HHG is that laser-driven electron-hole pairs evolve within a fluctuating many-body background. 
Many-body scattering processes encountered along the driven electron-hole trajectories modify their energy, coherence, recombination dynamics, and ultimately the emitted high-harmonic radiation~\cite{liu2016thin, silva2018correlated, lee2024many}. 
Previously, scattering from other charge carriers, coherent phonons~\cite{wang2017roles, rana2022high, rana2022probing, PhysRevLett.133.156901, zhang2024high, zhang2024enhanced, neufeld2022probing}, and long-range spin orders have been shown to modify high-harmonic spectra substantially~\cite{uchida2022high, murakami2022anomalous}.
Exploiting this sensitivity, prior studies have demonstrated that high-harmonic emission is responsive to a wide range of phase transitions, including superconducting, magnetic, charge-density-wave, and metal-insulator transitions ~\cite{uchida2022high, murakami2022anomalous, Tyulnev2025High, Alcala2022Quantum, bionta2021tracking}. 

Despite these advances, one universal and unavoidable interaction remains largely unexplored in the strong-field regime: scattering from thermally fluctuating lattice vibrations.
%Among the various many-body interactions in solids, electron (or hole) scattering from a thermally fluctuating lattice is of fundamental importance. 
At finite temperatures, individual atoms in the crystal lattice experience thermal fluctuations, resulting in a stochastically varying energy landscape that scatters carriers. 
This ubiquitous process governs charge transport in solids~\cite{fratini2020charge}, yet its role in strongly driven coherent electron-hole motion remains poorly understood. 
Thermal lattice fluctuations are expected to influence solid-state HHG due to their comparable length scales. 
This influence becomes most pronounced when the spatial extent of lattice fluctuations is shorter than the electron-hole trajectory length. 
In solid-state HHG, laser-driven electron-hole pairs typically traverse multiple unit cells~\cite{yue2020imperfect, Ghimire2019Solids}, implying that lattice fluctuations on the scale of a few unit cells can induce substantial scattering effects. 

Although atomic-scale lattice fluctuations are ubiquitous under thermal equilibrium, their intrinsic effects on HHG have not been explicitly demonstrated. This is largely owing to the lack of systematic control over thermal lattice fluctuations, including both their population and spatial correlation length. Additionally, strong lattice fluctuations are often intertwined with other phenomena, such as structural phase transitions, making it difficult to isolate their effect on HHG.

To address these challenges, we use Re$_6$Se$_8$Cl$_2$ superatomic crystals to explicitly unveil the effects of thermal lattice fluctuations on HHG. Superatomic crystals consist of coupled nanoclusters~\cite{zhong2018superatomic}, which lead to strong lattice fluctuations and well-defined low-energy optical phonons~\cite{lee2019hierarchical, li2020strong}. These phonons correspond to short-ranged cluster vibrations at distances comparable to or smaller than the typical electron-hole trajectories responsible for HHG~\cite{lee2019hierarchical}. Compared to conventional semiconductors, lattice fluctuations in superatomic crystals are more drastically modified by temperature~\cite{li2020strong}. The combination of strong lattice fluctuations, temperature tunability, and the lack of phase transitions in Re$_6$Se$_8$Cl$_2$ makes it an ideal platform to examine the interplay between HHG and thermally induced lattice disorder. We experimentally demonstrate that HHG is strongly influenced by thermal lattice fluctuations by measuring the temperature-dependent high-harmonic yield in Re$_6$Se$_8$Cl$_2$. We observe an abrupt enhancement of harmonic emission at low temperatures, where lattice fluctuations are suppressed. By comparing the experimental observations with a theoretical model, we extract the temperature-dependent dephasing time associated with carrier scattering from thermal lattice fluctuations. 

Our results validate using HHG to probe intrinsic lattice disorder without changing long-range order, symmetry breaking, or phase transitions. This spectral sensitivity can be further combined with HHG-based band structure and charge distribution reconstruction~\cite{vampa2015all, lakhotia2020laser}, enabling ultrafast, all-optical probing of carrier-phonon couplings under strong-field driving conditions. From a material science perspective, superatomic crystals exhibit widely tunable many-body interactions through atomically precise modification of cluster building blocks. Demonstrating HHG on superatomic crystals offers new opportunities for material control of HHG and other strong-field processes, such as lightwave electronics and Floquet engineering~\cite{borsch2023lightwave}.

\section{Results and Discussion}
\begin{figure}
\centering
\includegraphics[width= 0.8\linewidth]{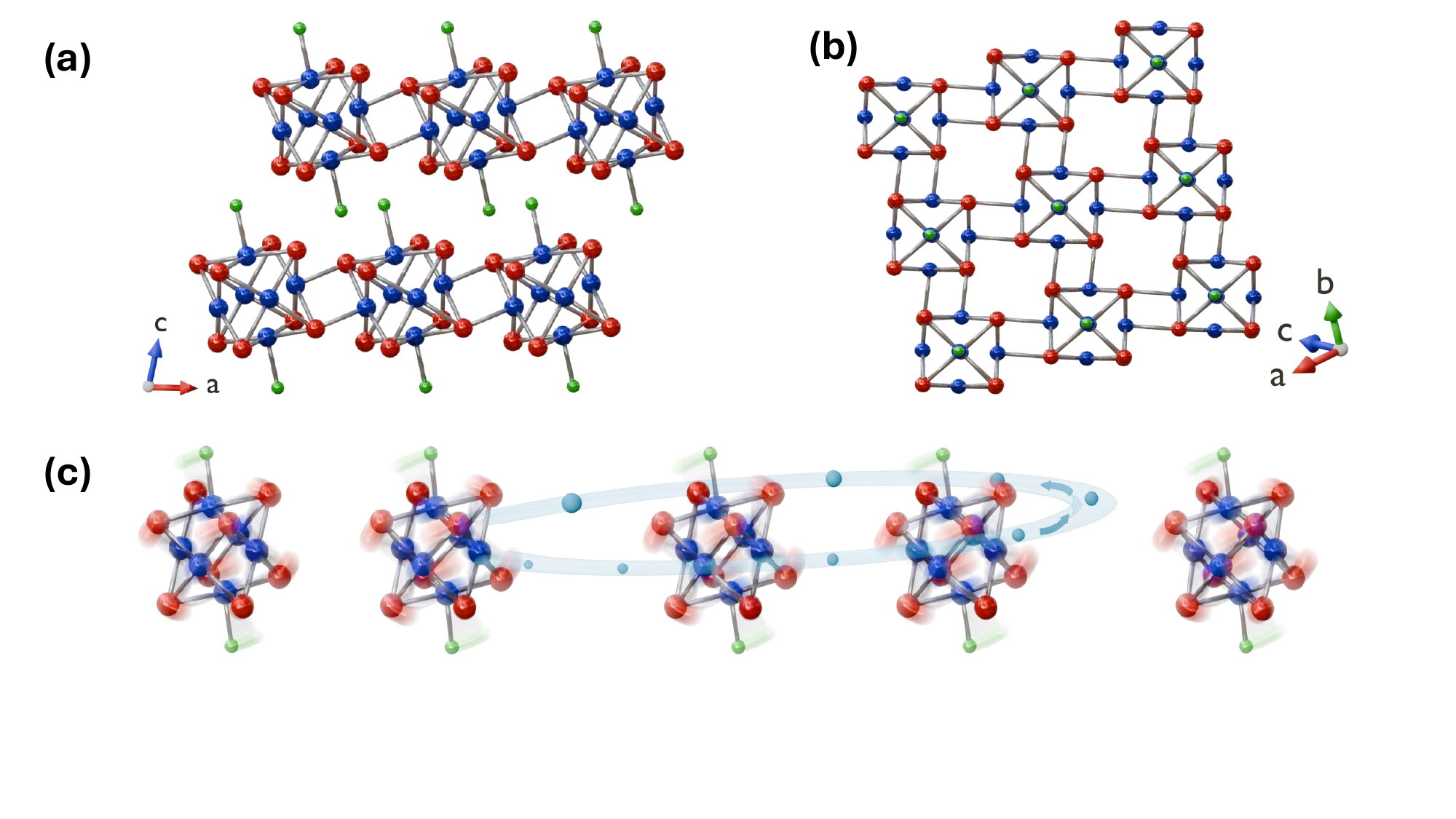}
\caption{\textbf{Crystal structure and thermal lattice fluctuations in Re$_6$Se$_8$Cl$_2$.} 
(a) Crystal structure viewed along the in-plane $b$ axis. 
(b) In-plane pseudo-square lattice geometry. 
(c) Thermal lattice fluctuations induce dynamic disorder, leading to scattering of laser-driven electron trajectories during the high-harmonic generation (HHG) process. 
%(D) At low temperatures, lattice fluctuations are suppressed, resulting in a more periodic lattice potential and reduced scattering.
%\textcolor{red}{Perhaps write explicit in figure file at C "inter+intra cluster phonons", and D "inter-cluster phonons". Why is intercluster at 7K when later it is mentioned that inter-cluster vanishes below 50 K (Fig.2b). Also for intracluster, why 280K instead of 336K? Perhaps add a vertical bar or axis showing the energies/frequencies of the phonon modes and the corresponding temperature, currently many different numbers to juggle from the text.}
}
~\label{schematic}
\end{figure}

 Re$_6$Se$_8$Cl$_2$ is a layered van der Waals semiconductor with an optical bandgap of approximately 1.5 eV~\cite{zhong2018superatomic}.
Figure~\ref{schematic}(a) illustrates the crystal structure, which consists of two-dimensional (2D) sheets formed by interconnected Re$_{6}$Se$_8$Cl$_2$ superatomic clusters. 
Each cluster consists of Re$_{6}$ octahedra encapsulated within an Se$_8$ cube. 
Within each plane, the clusters are covalently bonded to four nearest neighbors (Fig.~\ref{schematic}(b)), forming a pseudo square lattice. 
In the out-of-plane direction, each cluster is terminated by two Cl atoms. 
The resulting layers stack via van der Waals interactions to form a bulk crystal~\cite{choi2018two}. 

Owing to its lattice of interconnected superatomic clusters, Re$_6$Se$_8$Cl$_2$ exhibits a hierarchical phonon response that results in temperature-tunable lattice fluctuations involving collective cluster vibrations~\cite{lee2019hierarchical,li2020strong}. 
%Specifically, Re$_6$Se$_8$Cl$_2$ hosts high-frequency vibrational modes localized within individual clusters, as well as low-frequency intermolecular motion between neighboring clusters~\cite{lee2019hierarchical}. 
%These intra- and inter-cluster modes are well-separated in energy. 
%Ultrafast optical studies have shown that the high-frequency intra-cluster modes lie above 7 terahertz (THz), whereas the lower-frequency inter-cluster motion occurs near 2 THz. 
At room temperature, these phonon modes are thermally populated, leading to strong lattice fluctuations, as depicted in Fig.~\ref{schematic}(c). 
Microscopically, such structural fluctuations introduce dynamical disorder in the superatomic lattice, which can strongly scatter laser-driven electron-hole trajectories during high-harmonic generation. 
%Upon cooling, the thermal population of these modes is progressively suppressed, resulting in reduced lattice disorder (Fig.~\ref{schematic}(d)). 
%\sout{With decreasing sample temperature, the localized intra-cluster vibrations will be suppressed first, followed by the vanishing of inter-molecular vibrations from neighboring clusters at lower temperatures. }
As the temperature is lowered, the thermal population of phonons is progressively reduced, leading to a systematic suppression of lattice dynamics with enhanced long-range periodicity. 
During this cooling process, vibrational modes with larger characteristic energies are quenched first, followed by the depopulation of lower-energy modes at further reduced temperatures.
At temperatures well below the characteristic energy scale of the 
%\sout{inter-cluster modes, all optical phonons are depopulated, resulting in a crystal lattice with suppressed dynamical disorder. } 
lowest optical phonons, their occupation becomes negligible. 
The crystal thus approaches a regime in which phonon-driven lattice fluctuations are effectively extinguished, and dynamical disorder is strongly suppressed.

\begin{figure}
\centering
\includegraphics[width= 0.8 \linewidth]{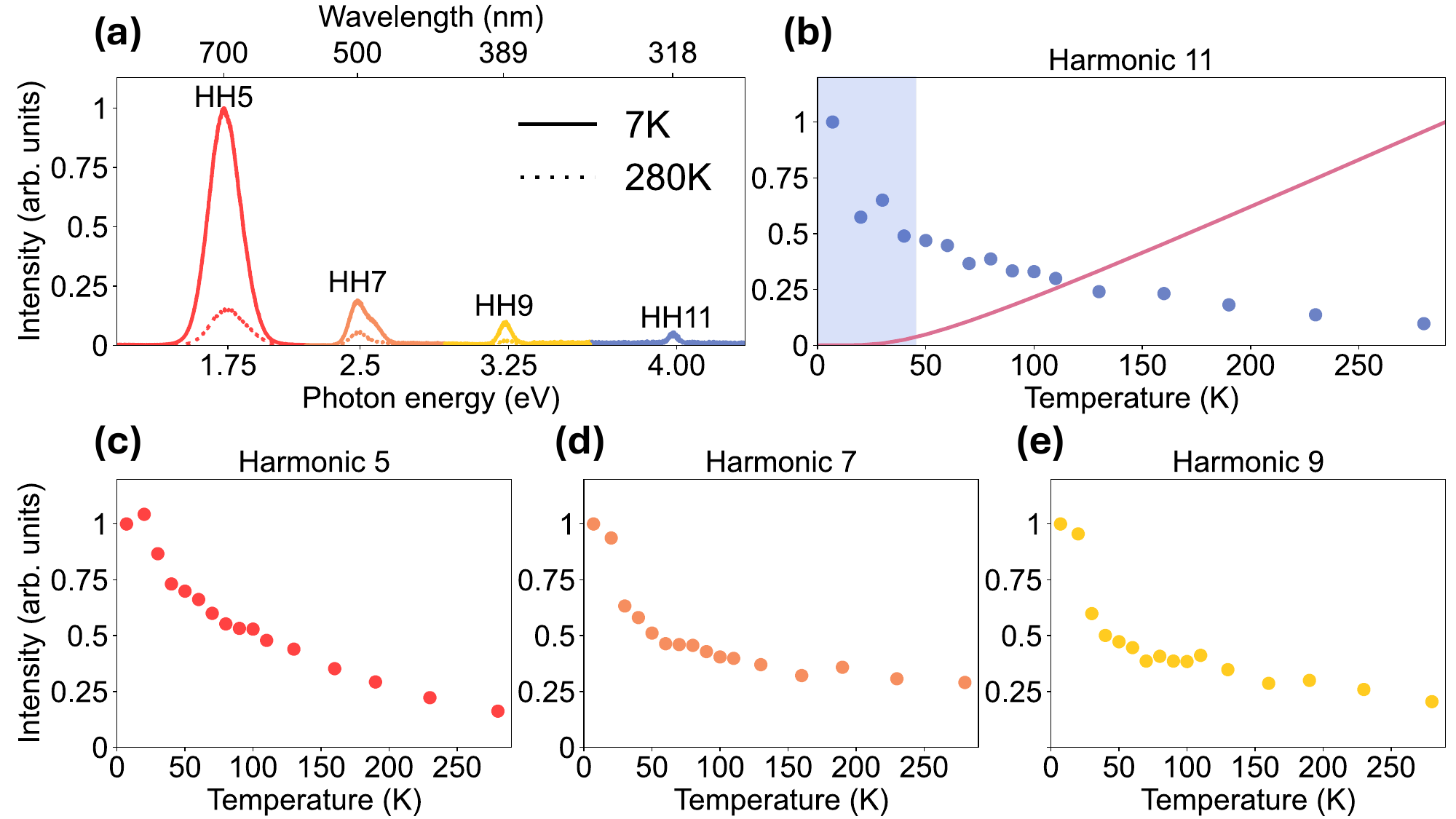}
\caption{\textbf{Dependence of HHG on temperature for Re$_6$Se$_8$Cl$_2$.} (a) Harmonic spectra of 5th to 11th harmonics recorded at 280 K (dotted line) and 7 K (solid line). (b) 11th harmonic yield recorded as a function of temperature. The red line depicts the estimated thermal population of a representative optical phonon at 2.6 THz. 
%\textcolor{red}{Where is the 2.6 THz mode taken from? address: There is a paper describing the 2.6 THz vibrational modes. We made this clear in the paragraph below.} 
The blue shaded area highlights the region in which the thermal phonon population has vanished, coinciding with the abrupt increase in harmonic yield. (c-e) 5th, 7th, and 9th, harmonic yields as a function of temperature, respectively.}
~\label{fig2}
\end{figure}

To investigate the influence of dynamical lattice fluctuations on the high-harmonic generation, we measure temperature-dependent high-harmonic spectra from bulk single crystal Re$_6$Se$_8$Cl$_2$. 
The sample is driven by a linearly polarized pulse with a duration of $\sim$85 fs. The wavelength is centered at 3.5 $\mu$m, corresponding to a photon energy of 0.35 eV, which is well below the bandgap of the sample. 
The laser polarization is aligned along the crystallographic a-axis (Fig.~\ref{schematic}(a)). 
The emitted radiation is measured in a reflection geometry. 
Representative high-harmonic spectra recorded at a driving intensity of 1.12 TW/cm$^2$ are shown in Fig.~\ref{fig2}(a). 
At 280 K, the 5$^{th}$, 7$^{th}$, and 9$^{th}$ harmonics are observed. Upon cooling to 7 K, the harmonic emission is strongly enhanced, accompanied by the strong intensification of the 11$^{th}$ harmonic. 
Figure~\ref{fig2}(b)-\ref{fig2}(e) displays the evolution of the high-harmonic yield as the sample temperature is reduced from 280 K to 7 K. 
The harmonic yield exhibits a nonlinear, multi-step increase with decreasing temperature. 
A gradual enhancement is observed between 280 K and 50 K, followed by an abrupt increase below 50 K that persists down to the lowest temperature measured. 
This behavior is common to all observed harmonics, with the relative enhancement becoming increasingly pronounced for higher-order harmonics. 
Also, the rate of increase in harmonic yield is different for each harmonic order as shown in Fig.~\ref{fig2}(b)-\ref{fig2}(d).
\newline
We attribute the nonlinear enhancement of high-harmonic emission at low temperatures to the suppression of dynamical lattice fluctuations, which reduces the scattering of laser-driven electron-hole pairs. This interpretation is corroborated by the abrupt enhancement of harmonic emission near 50 K, which coincides with the vanishing of 
%\sout{inter-cluster vibrational} 
the low-frequency optical phonon modes. These modes primarily consist of molecular vibrations arising from neighboring clusters. To further substantiate this connection, we plot the thermally-activated population of a representative optical phonon at 2.6 THz~\cite{lee2019hierarchical} as a function of temperature in Fig.~\ref{fig2}(b), estimated using a Bose–Einstein distribution. As the phonon population decreases, the harmonic intensity increases across all observed orders, with a sharp enhancement emerging as the 
%\sout{inter-cluster vibrations} 
low-frequency optical phonon mode becomes thermally depopulated.
% \colorbox{red}{\parbox{\linewidth}{Lun: vanish or population decrease below below 50 K? earlier mentioned inter-cluster mode was at 7K?}}. \textcolor{red}{Navdeep: As the temperature decreases, these modes gradually become thermally depopulated. In the literature, they are often called inter-cluster modes, but I have not plotted the eigenvectors in neighboring unit cells to confirm that classification. What I can say with confidence is that they are optical phonons. That is why I prefer to refer to them simply as optical phonon modes.} 

Notably, Re$_6$Se$_8$Cl$_2$ exhibits no structural or electronic phase transitions in this temperature range. Previous optical studies have reported a modest increase in the band gap upon cooling, which would instead be expected to suppress high-harmonic emission by reduced carrier tunneling~\cite{li2020strong}. Additionally, we measure the reflected driving pulse energy as a function of sample temperature, which does not show temperature dependence (Supplementary Fig. S6). These observations eliminate alternative explanations for the observed harmonic behavior based on electronic and nonlinear absorption effects. We therefore attribute the harmonic yield increase to reduced carrier scattering from lattice fluctuations.
%\sout{These results further suggest that, within the semi-classical picture, the laser-driven electron–hole trajectory length is comparable to or larger than the size of individual superatomic clusters. 
%Higher-order harmonics originate from carriers that acquire larger kinetic energies through prolonged laser acceleration and are thus more susceptible to dephasing from dynamical lattice disorder. 
%When the trajectory exceeds the cluster size, carriers scatter from neighboring clusters, making the high-harmonic response sensitive to inter-cluster lattice motion. 
%This interpretation is consistent with the observed harmonic-order dependence.}
%\colorbox{yellow}{\parbox{\linewidth}{As you also mentioned, the strike out part doesn't follow our reasoning. I don't know if this is necessary and if yes what should we write here.}}
%\colorbox{red}{Lun: why does it not follow the reasoning?}

%\sout{To corroborate these experimental observations and gain microscopic insight into the coupled electronic and lattice dynamics underlying the measured high-harmonic response, we adopt theoretical calculations in two steps.}
To understand the microscopic origin of the temperature-dependent high-harmonic response, we adopt theoretical calculations in two steps, to disentangle electronic and lattice contributions. The calculations are described in detail in the Methods section. 
%\sout{We combine first-principles electronic structure calculations with time-dependent simulations of the nonlinear light–matter interaction. }}
% Our theoretical analysis proceeds in two steps. 
Briefly, in the first step, first-principles density functional theory (DFT) calculations are performed for Re$_6$Se$_8$Cl$_2$ to calculate the ground-state band energies and momentum matrix elements. 
In the second step, these quantities are then used as input to time-dependent Semiconductor Bloch equation (SBE) simulations to capture the nonlinear light-matter interaction and HHG under strong-field excitation. The lattice dynamics are represented by a statistical ensemble of static distortions, obtained from symmetric positive and negative displacements of the relevant optical phonon modes. For each static distortion we re-calculate the band structure and the HHG spectrum, and then calculate the coherent average over all lattice configurations. 
%\sout{ This approximation is well justified because the characteristic phonon periods are significantly longer than the duration of the driving laser pulse, establishing a clear separation of electronic and  phononic time scales. 
%Furthermore, thermally populated phonon modes are incoherent, allowing the lattice to be treated as quasi-static during the ultrafast electronic evolution.}
This approximation is justified by the clear separation of timescales: the characteristic phonon periods are significantly longer than the duration of the driving laser optical period, so the electronic dynamics evolve in the presence of effectively frozen lattice coordinates. 
We also note that thermally populated phonons possess random phases, resulting in zero ensemble-averaged displacement but finite temperature-dependent fluctuations. 
%\textcolor{red}{Navdeep: The connection is through two separate points. Thermal phonons have random phases, so there is no coherent time-dependent lattice distortion, only a distribution of instantaneous displacements with zero mean and finite variance. Since their vibrational periods are much longer than the femtosecond electronic dynamics we simulate, the electrons effectively see a frozen lattice configuration. That is why we treat the lattice as quasi-static and then average over configurations.  Also, I have rephrased above two lines to add more clarity.}
The selected lattice configurations explicitly represent the temperature-dependent spatial distortions associated with the relevant optical phonon modes of Re$_6$Se$_8$Cl$_2$, thereby enabling a direct connection between lattice fluctuations and the nonlinear optical response.

In our calculations, high-harmonic spectra are generated for both equilibrium and phonon-distorted lattice configurations using a linearly polarized pulse along the crystallographic \textit{a}-axis, with a wavelength of $3.5\,\mu\mathrm{m}$, a peak electric field strength of $2.7 \times 10^{9}\,\mathrm{V/m}$, and a full width at half maximum (FWHM) of $85\,\mathrm{fs}$.
%peak intensity of 1 TW/cm$^{2}$ 
%\colorbox{red}{Do we know the refractive index?} {\color{blue} (Mette) For the calculation, the only thing that matters is the electric field strength and that is not impacted by the refractive index, it's only the intensity formula that has the refractive index in it. In my opinion we don't need to address this here} \colorbox{red}{Okay in that case, perhaps best to give the electric field strength number instead of peak intensity?} and a time period of 120 fs \colorbox{red}{Why longer than the experiment, or it this the field FWHM?}.
%\textcolor{red}{(Navdeep) This is the time period of the electric field. Should I write the time period of intensity?}
%\colorbox{red}{Yes better to be consistent with the experiment. Also, calling it time period is perhaps misleading, perhaps use "FWHM" or "intensity FWHM".}
%Convergence of the high-order harmonic spectra with respect to the number of electronic bands is systematically examined in the Supplemental Materiel (SM). 
%All  calculations shown below are performed with 42 valence and 14 conduction bands. \textcolor{red}{(Why? To capture the wave packet dynamics? Is this approach / including many bands equivalent to real space calculation? Could we add one sentence to explain the necessity for including many bands?) Answer: this is very standard when working in the velocity-gauge. I don't think we need to explain}
%\textcolor{red}{(I feel this paragraph is a bit too pedagogical. I tried simplified it. See if it makes sense.)}

In order to incorporate lattice degrees of freedom, we analyze the phonon modes of Re$_6$Se$_8$Cl$_2$. 
A primitive unit cell contains 16 atoms, giving rise to $48$ vibrational eigenmodes at each crystal momentum $\mathbf{k}$ in the Brillouin zone. 
These consist of three acoustic and 45 optical branches. 
%\sout{In principle, a complete finite-temperature treatment requires computing the full phonon dispersion and including contributions from all thermally populated modes throughout the Brillouin zone. }
%\sout{For each \textcolor{red}{momentum} $\mathbf{q}$\sout{-point}, the number of phonon branches remains 48, and 
The total number of lattice degrees of freedom is thus large and scales with the $\mathbf{k}$-point sampling. A 
rigorous evaluation of the temperature-dependent high-harmonic response would require sampling lattice displacements generated by all thermally occupied phonon modes across the whole Brillouin zone, with populations determined by Bose–Einstein statistics, which is computationally demanding due to the large unit cell involved. 
%\sout{However, computing the full phonon dispersion and constructing thermally averaged lattice ensembles for this cluster-based system is computationally demanding.}

To capture the dominant lattice effects while maintaining computational feasibility, we adopt a reduced model. 
We compute the phonon modes only at the Brillouin-zone center ($\Gamma$ point) using density functional perturbation theory, obtaining mode frequencies $\Omega_\nu$ and normalized eigenvectors $\mathbf{e}_{\alpha}^{(\nu)}$ for each atomic degree of freedom $\alpha$, where $\nu$ labels the phonon mode index. 
%running over all $3N$ vibrational modes of the unit cell (with $N$ atoms), including both acoustic and optical branches
Among these modes, we retain the four lowest-frequency optical phonons, which are expected to dominate in the temperature range around 60 K to 140 K. 
These modes are treated as thermally active, and their amplitudes are allowed to vary with temperature.
Lattice dynamics are sampled at sixteen snapshots over one period of the lowest-frequency optical phonon: eight equally spaced points in the first quarter of the period and eight equally spaced points in the third quarter, so that each snapshot represents a distinct instantaneous lattice configuration.
The phonon oscillation amplitudes are obtained from the thermal expectation value of the harmonic oscillator displacement, which reduces to the classical limit for $k_{B}T \gg \hbar \Omega_{\nu}$.
Thermal excitation of these modes induces time-dependent lattice displacements
\begin{equation}
\mathbf{u}_{\alpha}(t) = \mathbf{u}_{\alpha}^0 + \tilde{\mathbf{u}}_{\alpha}(t), \quad
\tilde{\mathbf{u}}_{\alpha}(t) = \mathrm{Re} \Bigg[ \sum_{\nu} \sqrt{\frac{2 k_B T}{M_{\alpha} \Omega_\nu^2}} \mathbf{e}_{\alpha}^{(\nu)} e^{-i \Omega_{\nu} t} \Bigg],
\end{equation}
where $M_\alpha$ is the atomic mass.
These lattice fluctuations modulate the electronic structure by inducing bandgap variations and modifying the dipole couplings, reflecting the underlying electron–phonon interactions~\cite{zhang2024high}.
For a given temperature $T$, we construct a set of instantaneous lattice configurations incorporating these displacements, and the corresponding ground-state electronic quantities are recalculated for each configuration. 
The resulting electronic states are then propagated in time using the SBEs under the presence of external driving field. 
The currents computed for multiple thermally perturbed lattice configurations at a given temperature are coherently summed to obtain an ensemble-averaged, time-dependent current, which reflects the influence of incoherent phonons on high-harmonic generation.
This framework establishes a direct link between thermal lattice fluctuations, electron–phonon interactions, and the temperature-dependent modification of the high-harmonic response.

\begin{figure}[htbp]
\centering
\includegraphics[width=0.8\linewidth]{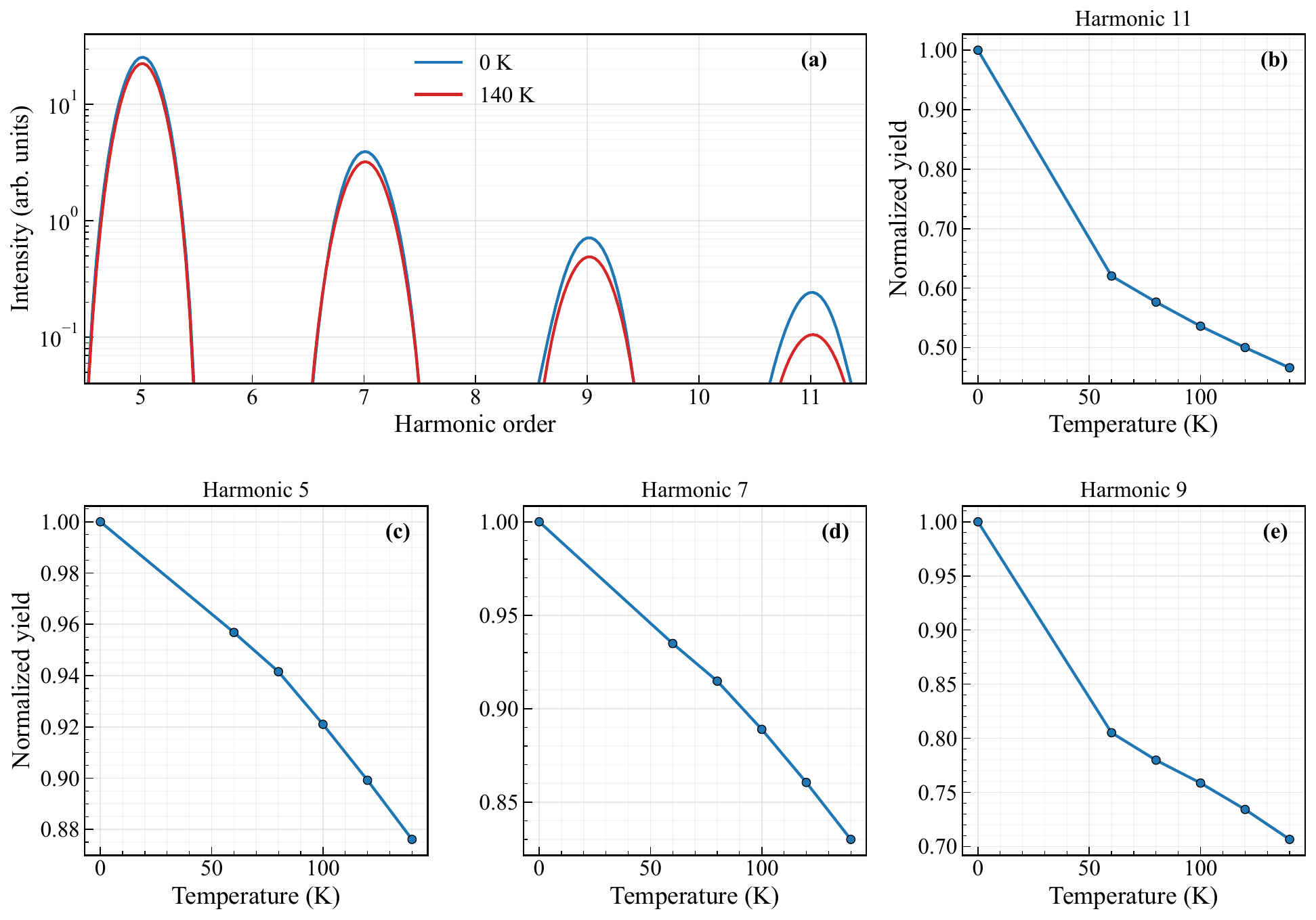}
\caption{
\textbf{Temperature-dependent high-harmonic response of Re$_6$Se$_8$Cl$_2$.}
(a) Theoretically calculated high-harmonic spectra of monolayer Re$_6$Se$_8$Cl$_2$ at 0~K (blue) and 140~K (red).
(b-e) Temperature dependence of the integrated high-harmonic yield for the 11th (b), 5th (c), 7th (d), and 9th (e) harmonics.
All yields are normalized to their respective maximum values.
}
\label{fig3}
\end{figure}
%\textcolor{red}{Would it be helpful to show the calculated phonon spectrum?} 
%In all theoretical results presented here, the first four optical phonon modes at the $\Gamma$-point are included. 
%For the lowest-frequency optical mode, sixteen distinct time snapshots distributed over one phonon period are considered to sample the lattice dynamics.
%This mode has a frequency of approximately 1.6~THz.
%, corresponding to a thermal activation threshold of about 80~K. 
%Calculations are performed for temperatures between 60~K and 140~K, spanning the regime where these optical phonons become thermally populated.

Figure~\ref{fig3} shows the temperature dependence of the high-harmonic response computed using the framework described above.
The ensemble averaging over thermally perturbed lattice configurations effectively captures the effect of incoherent phonons at finite temperatures. 
%This approach is justified because the phonon period are much longer than the driving laser pulse duration, and the thermally populated phonons are incoherent allowing the lattice to be treated as quasi-static within the Born-Oppenheimer approximation during the electronic dynamics.
The high-harmonic spectra at 0 K (blue) and at 140 K (including phonon perturbations, red) are presented in Fig.~\ref{fig3}(a). 
As in the experimental results shown in Fig.~\ref{fig2}(a), only odd-order harmonics are observed, and the relative intensities exhibit good agreement with the measured values. 
%\colorbox{red}{Lun: nice. Curious how this looks in linear scale?} 
%\textcolor{red}{I tried to plot this in the linear scale. I doesn't look as good as in the experiments.}
The overall harmonic intensity is reduced at 140 K, with the suppression becoming more pronounced at the higher harmonic orders. 
%This trend reflects the increased sensitivity of higher-order harmonics to the phase accumulated over longer electron-hole trajectories associated with higher harmonics.
%As the temperature increases, a greater number of phonon modes become thermally populated, resulting in larger atomic displacements from equilibrium and increased structural disorder in the lattice. 
As the temperature increases, atomic displacements from equilibrium grow, leading to enhanced structural disorder in the lattice.
These displacements modulate local hopping amplitudes and band energies, introducing spatially and temporally fluctuating phase shifts in the driven electronic polarization~\cite{rana2022high}.
This growing disorder progressively suppresses the overall high-harmonic response.
To quantify the role of these fluctuations, we compare coherent and incoherent ensemble summations over thermally distorted configurations (See supplemental Fig.~S4). The coherent sum exhibits a stronger reduction in harmonic yield, indicating that configuration-dependent phase variations introduce partial destructive interference. In contrast, the incoherent sum shows a weaker decrease, reflecting primarily the diminished emission strength of each distorted geometry.

Figure~\ref{fig3}(b)–\ref{fig3}(e) shows the temperature dependence of the integrated yield for harmonics 11, 5, 7, and 9, respectively.
The integrated yield of each harmonic decreases with increasing temperature. The rate of this decrease is harmonic-dependent and becomes more pronounced at higher harmonic orders, in agreement with the experimental behavior. This trend reflects the increased sensitivity of higher-order harmonics to the phase accumulated over longer electron-hole trajectories associated with higher harmonics.

Previously, harmonic order-dependent intensity reduction has been observed in photoexcited semiconductors. The signal reduction was attributed to incoherent scattering from photo-injected hot carriers, which can be modeled by incorporating an effective dephasing time in the SBE for electron-hole pairs~\cite{heide2022probing, rana2022generation}.
%\textcolor{red}{Our analysis indicates that the temperature-induced suppression originates from two concurrent mechanisms: a reduction in the harmonic emission amplitude from each distorted configuration due to lattice-induced electronic modifications, and an additional attenuation arising from phase mismatch in the coherent ensemble average. The relative contribution of these two effects governs the overall yield reduction.}
Our results suggest that incoherent phonons act as an effective dephasing channel for the driven electron-hole polarization.
As the temperature increases, thermally populated phonons distort the lattice and induce fluctuations, which reduce the phase coherence of the driven electronic motion. 
Within this picture, phonon-induced fluctuations enter as stochastic phase noise in the interband polarization, analogous to a temperature-dependent reduction of the electronic dephasing time $T_{2}$.
This dephasing weakens the constructive interference required for efficient high-harmonic emission, leading to a reduced yield. 
Because different harmonics probe different regions of the band structure and are generated over different temporal windows, their susceptibility to phonon-induced disorder varies with harmonic order, leading to distinct suppression rates. 
Higher-order harmonics are preferentially generated later in the optical cycle and involve larger band curvature and excursion, making them more susceptible to phonon-induced fluctuations~\cite{dudovich2006birth, vampa2015linking}.

\begin{figure}[htbp]
\centering
\includegraphics[width=0.8\linewidth]{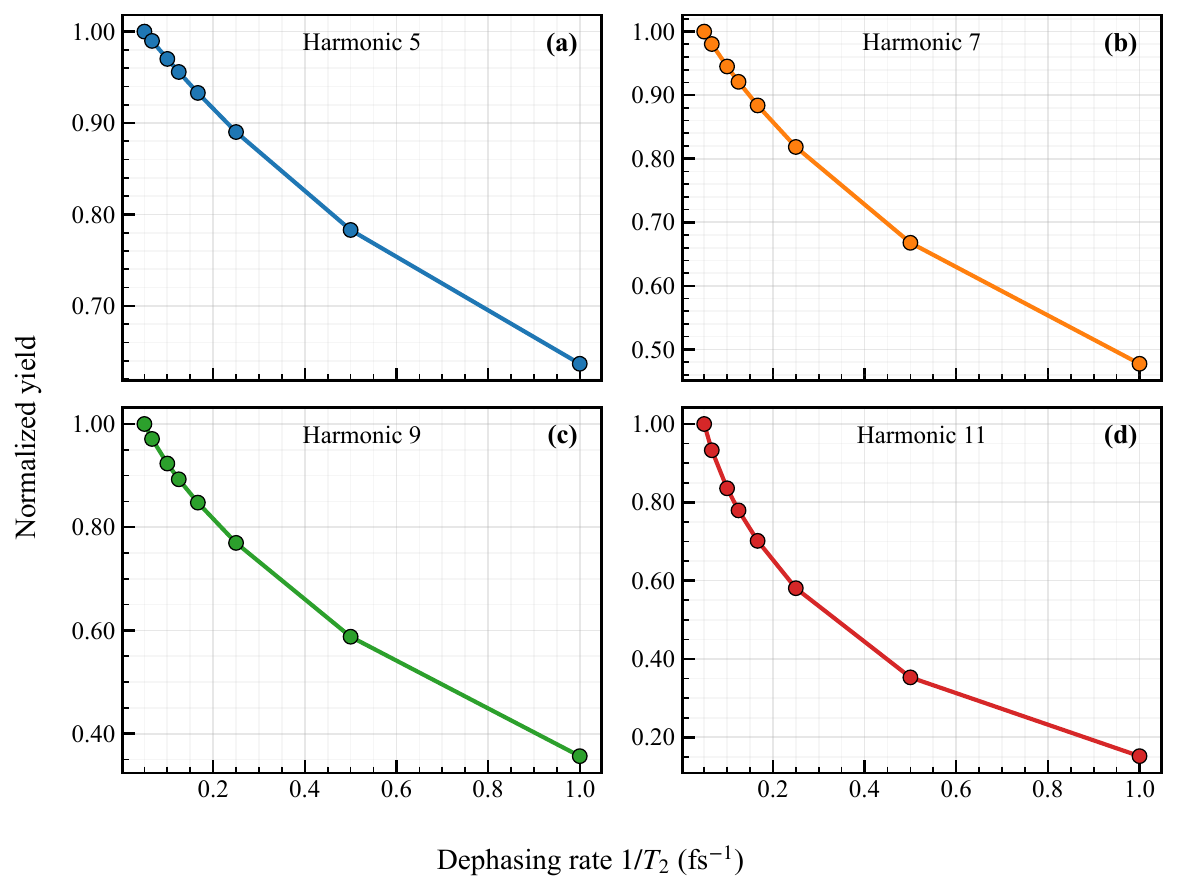}
\caption{
\textbf{Sensitivity of high-harmonic generation to electronic dephasing in Re$_6$Se$_8$Cl$_2$.}
Normalized integrated high-harmonic yield as a function of dephasing rate $1/T_{2}$ for the 5th (a), 7th (b), 9th (c), and 11th (d) harmonics.
All yields are normalized to their respective maximum values.
}
\label{fig4}
\end{figure}
The agreement between experiment and first-principles modeling suggests that phonon-induced electronic decoherence governs the temperature-dependent suppression of the harmonic yield. The decoherence can be characterized by a temperature-dependent effective dephasing rate in SBE. %\sout{These results motivate a description of the temperature dependence in terms of an effective dephasing rate.}
To further substantiate this interpretation, we perform a series of HHG calculations in undistorted lattices with varying dephasing time $T_{2}$.
%\sout{To further substantiate this interpretation, we perform a series of calculations of HHG in the spatial equilibrium configuration, in which we vary the dephasing time characterized by $T_{2}$.}
%\sout{The dephasing rate induced by optical phonons follows the Bose-Einstein occupation, reflecting that phonon absorption and emission drive fluctuations of the electronic states. 
%Accordingly, the electronic dephasing time is expressed as \cite{krummheuer2002theory}
%$T_2 \propto \dfrac{1}{n_{BE}(\Omega_{\rm op}, T )} = e^{\hbar \Omega_{\rm op}/k_B T}-1$, 
%\colorbox{red}{There should be a typo in this Eq.} \textcolor{red}{$\hbar$ factor was missing.}
%where $\Omega_{\rm op} = 2.76$ THz denotes the optical phonon frequency.}
Increasing the temperature enhances scattering processes that disrupt electronic coherence, leading to a larger dephasing rate $1/T_{2}$ and consequently a shorter coherence time. 
Figure~\ref{fig4} shows the integrated odd harmonic yield (5$^{th}$–11$^{th}$) as the electronic dephasing time is varied in Re$_6$Se$_8$Cl$_2$ for equilibrium configuration.
The harmonic yield decreases as the dephasing time is reduced, demonstrating that longer electronic coherence times enhance the efficiency of high-order harmonic generation. 
The decrease is more pronounced for higher-order harmonics, consistent with both the experimental observations and the \textit{ab initio} phonon model.

This controlled variation establishes a direct correspondence between temperature-induced lattice fluctuations and an effective reduction of the electronic coherence time inferred from Fig.~\ref{fig3}.
%\sout{For the 11$^{th}$ harmonic, the yield is reduced to approximately 30\% of its equilibrium value at 140 K, while the corresponding reduction is about 45\% in the incoherent phonon model and 25\% in the experimental data.} 
As an example, we highlight the comparison between the yield of 11th harmonic as a function of temperature (Fig. 2b) and dephasing time (Fig. 4d). As the temperature increases from 7 K to 140 K, the experimental harmonic yield reduces to 25\% of its peak value. Based on calculation (Fig. 4d), this signal drop corresponds to an effective dephasing rate at 0.75 fs$^{-1}$, or a equivalent dephasing time at 1.33 fs. Meanwhile, the dephasing time drastically increases when the temperature is 10 K, reaching $\sim$ 20 fs for 11th harmonic.
%In particular, low-frequency acoustic modes and mode-dependent electron-phonon couplings, which contributes additional phase noise, are not fully captured in the present implementations, and their inclusion is expected to further improve quantitative agreement with temperature-dependent measurements.

The quantification of a thermal lattice fluctuation-induced dephasing time has several important implications. Historically, theoretical descriptions of HHG in solids often need to introduce a phenomenological dephasing time on the order of $\sim$ 1 fs to match experimental results~\cite{vampa2015semiclassical, vampa2014theoretical, yue2020imperfect}. Such a short dephasing time is often attributed to electron-electron scattering~\cite{vampa2015semiclassical, vampa2014theoretical, heide2022probing}. However, the physical origin of this dephasing time is poorly understood. More importantly, it is unclear whether the introduced dephasing term is physically inherent or is a mere convenience for theoretical descriptions. Our results indicate that for materials with strong electron-phonon couplings, thermal lattice fluctuations can be the dominant electronic dephasing source, leading to a few-femtosecond dephasing times even at moderately low temperatures. Meanwhile, an order-of-magnitude increase in dephasing time is estimated when thermal fluctuations are suppressed at low temperatures. Importantly, the $>$ 10 fs dephasing time at low temperatures could be readily measured in real time using state-of-the-art ultrafast laser technologies. More generally, the demonstration of HHG on structurally tunable superatomic crystals allows material engineering of strong-field phenomena under various many-body fluctuations through atomically precise cluster synthesis.

%More generally, while this work highlights a superatomic crystal as a test material due to its strong electron-phonon couplings, the framework can be extended to other systems with many-body fluctuations.

\backmatter
\section{Methods}
\subsection{Experimental methods}
We performed temperature-dependent high-harmonic generation on Re$_6$Se$_8$Cl$_2$ single crystals. Re$_6$Se$_8$Cl$_2$ single crystals were grown using chemical vapor transport methods, which has been described in previous publications~\cite{zhong2018superatomic, teleford2018doping}. The resulting single crystals exhibit smooth, milimeter-sized surfaces, which are ideal for optical studies. The incident laser beam was aligned perpendicular to the crystal surface and thus the 2D sheets formed by interconnected Re$_6$Se$_8$Cl$_2$ superatomic clusters. The crystal phase and crystallographic orientation was confirmed by single-crystal X-ray diffraction using a Bruker APEX2 diffractometer with Mo K$_{\alpha}$ radiation. High-harmonic generation was achieved by a pump pulse centered at 3.5 µm. This pulse is obtained by difference-frequency generation of the signal and idler outputs of an optical parametric amplifier, pumped by a Ti:Sapphire laser operating at a 1kHz repetition rate. The 3.5 µm pulse was focused on a single crystal of Re$_6$Se$_8$Cl$_2$ by a 100 mm focal length CaF$_2$ lens. The pump power was controlled by two wire-grid polarizers (Thorlabs WP25M-IRA). The polarization of the pump pulse was controlled by a MgF$_2$ half-wave plate. The sample was mounted in a closed-loop cryostat (Advanced Research Systems), allowing for temperature control between 7 K to 280 K. The high-harmonic radiation was collected in a reflection geometry and focused into a spectrometer (Princeton Instrument SP-2300i). The high-harmonic spectra were recorded by an electron multiplying charge-coupled device (EMCCD) detector (Princeton Instruments ProEM+ 1600$^2$). The reported harmonic spectra were corrected to account for wavelength-dependent detection efficiencies.
\subsection{Harmonic-generation calculations}
The ground-state electronic properties of monolayer Re$_6$Se$_8$Cl$_2$ are computed using the Perdew-Burke-Ernzerhof (PBE) exchange-correlation functional together with fully relativistic pseudopotentials including spin-orbit couplings using Quantum Espresso~\cite{giannozzi2009quantum}. 
Plane-wave kinetic energy cutoffs of 110~Ry for the wavefunctions and a $11\times11\times1$ Monkhorst-Pack \emph{k}-point mesh are used to sample the Brillouin zone. 
As shown in Supplementary Fig.~S1, the material exhibits an indirect bandgap with the valence band maximum at $\Gamma$ and conduction band minimum at the $\mathsf{T}$ point. 

The electron dynamics are  described by the time evolution of the density matrix elements $\rho^{\mathbf{k}}_{nm}(t)$  at crystal momentum $\mathbf{k}$ for bands $n$ and $m$, with the density matrix operator defined from the time-dependent wave function $|\Psi(t)\rangle$ as $\hat{\rho}(t)=|\Psi(t)\rangle \langle \Psi(t) |$.
%,  obtained by solving the SBEs.
%\textcolor{blue}{The density matrix is defined as
%\begin{equation}
%\rho^{\mathbf{k}}_{nm}(t)=\langle a^{\dagger}_{m\mathbf{k}}(t)a_{n\mathbf{k}}(t)\rangle ,
%\end{equation}
%where $a^{\dagger}_{n\mathbf{k}}$ and $a_{n\mathbf{k}}$ are creation and annihilation operators for an electron in band $n$ at crystal momentum $\mathbf{k}$. }

For a spatially uniform driving field characterized by a vector potential $\mathbf{A}(t)$, the equations of motion for the density matrix elements (equivalent to the semiconductor Bloch equations) reads
\begin{equation}
 i \dot{\rho}^{\mathbf{k}}_{nm}(t)
= \omega^{\mathbf{k}}_{nm}\,\rho^{\mathbf{k}}_{nm}(t)
+ \mathbf{A}(t) \cdot \sum_l \left[
  \mathbf{p}^{\mathbf{k}}_{nl}\,\rho^{\mathbf{k}}_{lm}(t)
- \rho^{\mathbf{k}}_{nl}(t) \mathbf{p}^{\mathbf{k}}_{lm}
  \right],
\end{equation}
where $\omega_{nm}^{\mathbf{k}} = \varepsilon_n^{\mathbf{k}} - \varepsilon_m^{\mathbf{k}}$ are the band energy difference at crystal momentum ${\mathbf{k}}$, and $\mathbf{p}_{nm}(\mathbf{k})$ are the momentum matrix elements extracted from the electronic structure calculation. For more details, see \cite{yue2022introduction}. 
The dynamics are computed in the velocity gauge within the dipole approximation. 
A phenomenological dephasing time $T_{2}=10\,\mathrm{fs}$ is included to account for residual scattering. 
The dephasing term is implemented in the field-dressed basis, as described in further detail in Ref.~\cite{yue2022introduction}.
%\textcolor{red}{(How is T2 included in the calculation? Should there be a T2 term in the above equation?)}
The microscopic current is obtained as
\begin{equation}
\mathbf{j}(t) = - \mathsf{Tr}\{[\hat{\mathbf{p}} + \mathbf{A}(t)]\hat{\rho}\}
~=~ N^{-1} \sum_{\mathbf{k}\in \mathrm{BZ}} \sum_{mn}
\left[
\mathbf{p}^{\mathbf{k}}_{mn}
+ \delta_{mn}\mathbf{A}(t)
\right]
\rho^{\mathbf{k}}_{nm}(t),
\end{equation}
where $N$ denotes the total number of $\mathbf{k}$ points used to sample the Brillouin zone, and $\delta_{mn}$ is the Kronecker delta.

%\textcolor{red}{(How is the trace defined?)}
The high-harmonic spectrum is computed as 
\begin{equation}
S(\omega) = \omega^{2} \left| \frac{1}{\sqrt{2\pi}} \int_{-\infty}^{\infty} \mathbf{j}(t) \, e^{i \omega t} dt \right|^{2}.
\end{equation}
Convergence of the high-order harmonic spectra with respect to the number of electronic bands is systematically examined in the Supplemental Materiel (SM). 
All  calculations shown here are performed with 42 valence and 14 conduction bands.
%\bmhead{Supplementary information}

\section{Acknowledgment}
L.H. and H.L. acknowledge support from the National Science Foundation under award number 2325212. Work at Louisiana State University (LSU) was supported by the National Science Foundation under grant number PHY-2409463, and high-performance computational resources were provided by LSU and the Louisiana Optical Network Infrastructure (LONI). L.Y. acknowledges startup funds from Binghamton University. B.Z., Z.Y. and H.W. acknowledge the support for single-crystal XRD by the U.S. Department of Energy, Office of Science, Basic Energy Sciences, Materials Sciences and Engineering Division. Research on Superatomic Materials at Columbia University is supported by the AFOSR MURI on Programmable Transport in Superatomic Materials under award number FA9550-25-1-0288. L.H. acknowledges support from the NSF Graduate Research Fellowship Program.

\bibliography{solid_HHG}
\end{document}

\title[Article Title]{Supplemental Material: Probing lattice fluctuations using solid-state high-harmonic spectroscopy }

\author[1]{\fnm{Lance} \sur{Hatch}}
\equalcont{These authors contributed equally to this work.}
\author[2]{\fnm{Navdeep} \sur{Rana}}
\equalcont{These authors contributed equally to this work.}
\author[3]{\fnm{Shoushou} \sur{He}}
\author[3]{\fnm{Jessica} \sur{Yu}}
\author[4]{\fnm{Boyang} \sur{Zhao}}
\author[4]{\fnm{Yu} \sur{Zhang}}
\author[4]{\fnm{Haidan} \sur{Wen}}
\author[3]{\fnm{Xavier} \sur{Roy}}
\author[5]{\fnm{Lun} \sur{Yue}}
\author[2]{\fnm{Mette} \sur{Gaarde}}
\author[1]{\fnm{Hanzhe} \sur{Liu}}

\affil[1]{\orgdiv{Department of Chemistry}\orgname{, Purdue University}\orgaddress{\city{, West Lafayette}, \postcode{47907}, \state{Indiana}, \country{USA}}}
\affil[2]{\orgdiv{Department of Physics and Astronomy}, \orgname{Louisiana State University}, \orgaddress{\city{Baton Rouge}, \postcode{70803}, \state{Louisiana}, \country{USA}}}
\affil[3]{\orgdiv{Department of Chemistry}, \orgname{Columbia University}, \orgaddress{\city{New York}, \postcode{10027}, \state{New York}, \country{USA}}}
\affil[4]{\orgdiv{Materials Science Division}, \orgname{Argonne National Laboratory}, \orgaddress{\city{Lemont}, \postcode{60439}, \state{Illinois}, \country{USA}}}
\affil[5]{\orgdiv{Department of Physics}, \orgname{Binghamton University, State University of New York}, \orgaddress{\city{Vestal}, \postcode{13902}, \state{New York}, \country{USA}}}

%\pacs{}

%%%%%%%%%%%%%%%%% END OF PREAMBLE %%%%%%%%%%%%%%%%

\maketitle 
 
\section{\texorpdfstring
  {Band structure of monolayer Re$_6$Se$_8$Cl$_2$}
  {Band structure of monolayer Re6Se8Cl2}}

\begin{figure}[h!]
\centering
\includegraphics[width=0.8\linewidth]{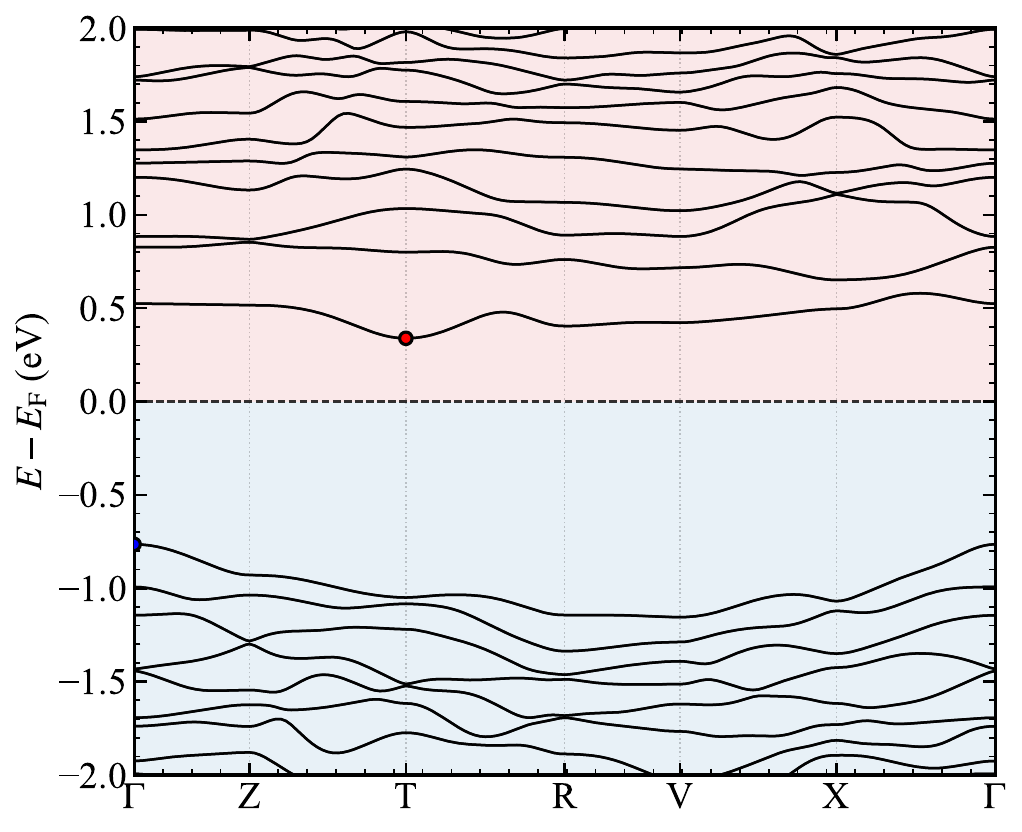}
\caption{Calculated DFT band structure of monolayer Re$_6$Se$_8$Cl$_2$ using PBE-SOC.} 
\label{bandstructure}
\end{figure}

Figure~S\ref{bandstructure} presents the band structure of monolayer Re$_6$Se$_8$Cl$_2$ obtained using density functional theory (DFT) as implemented in the Quantum ESPRESSO package. 
The fermi level is marked with dashed line at 0~eV.
Monolayer Re$_6$Se$_8$Cl$_2$ has an indirect bandgap of 1.1 eV, with the valence band maximum at $\Gamma$ and conduction band minimum at the $\mathsf{T}$ point.
\renewcommand{\figurename}{Supplementary Fig.}

\section{Convergence of HHG Spectra with number of bands}
\renewcommand{\figurename}{Supplementary Fig.}
\begin{figure}[h!]
\centering
\includegraphics[width=0.8\linewidth]{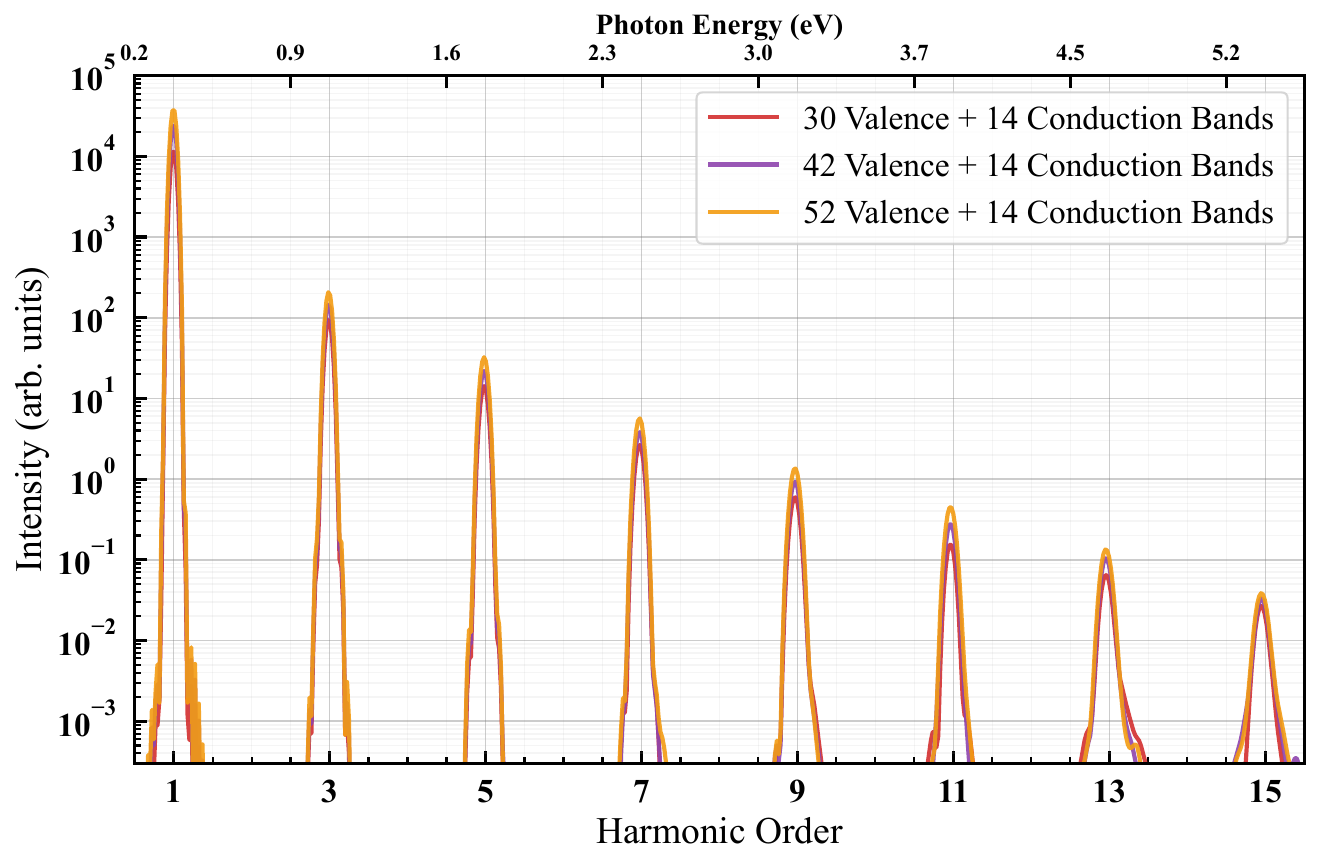}
\caption{Convergence of calculated high-order harmonic spectra as a function of the number of valence and conduction bands, in the absence of phonon dynamics. } 
\label{BandConvergence}
\end{figure}
As shown in the Fig.~S\ref{BandConvergence}, the harmonic response approach convergence upon increasing the band manifold beyond a critical number. 
Phonon dynamics are excluded in the convergence test, allowing the purely electronic response to be isolated.

\section{Modeling of thermal lattice fluctuations}
Thermal lattice fluctuations are modeled by displacing atoms along the phonon normal modes obtained from density-functional perturbation theory calculations using the phonon module of Quantum ESPRESSO~\cite{giannozzi2009quantum,giannozzi2017advanced}. The phonon eigenvectors $e_{\alpha}^{(\nu)}$ and frequencies $\Omega_\nu$ are computed at the $\Gamma$ point, where $\nu$ labels the phonon mode and $\alpha$ denotes the atomic cartesian coordinate.

\subsection{Classical Thermal Lattice Fluctuations}

We consider a system of $N$ atoms described by the classical Hamiltonian
\begin{equation}
H = T + V ,
\end{equation}
where the kinetic energy is
\begin{equation}
T = \frac{1}{2}\sum_{\alpha} M_\alpha \dot{\mathbf{u}}_\alpha^2 ,
\end{equation}
and $V$ is the potential energy of the lattice. Here $\mathbf{u}_\alpha = \mathbf{R}_\alpha - \mathbf{R}_\alpha^0$ denotes the displacement of atom $\alpha$ from its equilibrium position $\mathbf{R}_\alpha^0$.

For small displacements the potential energy can be expanded within the harmonic approximation,
\begin{equation}
V = V_0 + \frac{1}{2}\sum_{\alpha\beta} \mathbf{u}_\alpha D_{\alpha\beta} \mathbf{u}_\beta ,
\end{equation}
where $D_{\alpha\beta}$ is the dynamical matrix evaluated at equilibrium. Diagonalization of the dynamical matrix yields a set of normal modes with frequencies $\Omega_\nu$ and eigenvectors $\mathbf{e}_\alpha^{(\nu)}$.

The lattice displacement can therefore be expressed as a superposition of phonon modes,
\begin{equation}
\mathbf{u}_\alpha(t) =
\mathbf{u}_\alpha^0 +
\mathrm{Re}
\left[
\sum_{\nu} A_\nu
\mathbf{e}_\alpha^{(\nu)}
e^{-i\Omega_\nu t}
\right].
\end{equation}

Within the classical harmonic approximation, each phonon mode behaves as an independent harmonic oscillator. In thermal equilibrium the equipartition theorem assigns an average energy $k_B T$ to each mode, which determines the mode amplitude
\begin{equation}
A_\nu =
\sqrt{\frac{2k_B T}{M_\alpha \Omega_\nu^2}} .
\end{equation}

The thermally induced lattice displacement therefore takes the form
\begin{equation}
\mathbf{u}_{\alpha}(t) = \mathbf{u}_{\alpha}^0 + \tilde{\mathbf{u}}_{\alpha}(t), \quad
\tilde{\mathbf{u}}_{\alpha}(t) =
\mathrm{Re}\Bigg[
\sum_{\nu}
\sqrt{\frac{2 k_B T}{M_{\alpha} \Omega_\nu^2}}
\mathbf{e}_{\alpha}^{(\nu)}
e^{-i \Omega_{\nu} t}
\Bigg],
\end{equation}
where $M_\alpha$ is the atomic mass.

\subsection{Quantum Lattice Fluctuations}

In the quantum description, each normal mode of the lattice corresponds to a quantum harmonic oscillator with discrete energy levels
\begin{equation}
E_m = \hbar \Omega_\nu \left(m+\frac{1}{2}\right),
\end{equation}
where $\Omega_\nu$ is the phonon frequency of mode $\nu$. At finite temperature, the occupation of these levels follows the Bose--Einstein distribution
\begin{equation}
n_B(\Omega_\nu,T) =
\frac{1}{e^{\hbar\Omega_\nu/k_B T}-1}.
\end{equation}

The thermal expectation value of the mean-square displacement for a phonon mode is therefore
\begin{equation}
\langle |A_\nu|^2 \rangle =
\frac{\hbar}{2M_\alpha \Omega_\nu}
\left(2n_B(\Omega_\nu,T)+1\right),
\end{equation}
where the additional $+1$ term accounts for the zero-point motion of the lattice that persists even at zero temperature.

Using this temperature-dependent amplitude, the lattice displacement can again be written as a superposition of phonon modes,
\begin{equation}
\tilde{\mathbf{u}}_{\alpha}(t) =
\mathrm{Re}\Bigg[
\sum_{\nu}
\sqrt{
\frac{\hbar}{2M_\alpha \Omega_\nu}
\left(2n_B(\Omega_\nu,T)+1\right)
}
\mathbf{e}_{\alpha}^{(\nu)}
e^{-i\Omega_\nu t}
\Bigg].
\end{equation}

To examine the impact of quantum lattice fluctuations on the high-harmonic response, we compute thermally distorted lattice configurations using the quantum displacement amplitude derived above. 
In this analysis we include the same four optical phonon modes considered in the classical calculations. 
The amplitude of each mode is determined by the Bose-Einstein occupation factor $n_B(\Omega_\nu,T)$, which increases with temperature and therefore enhances the magnitude of the lattice displacements.

\begin{figure}[h!]
\centering
\includegraphics[width=0.8\linewidth]{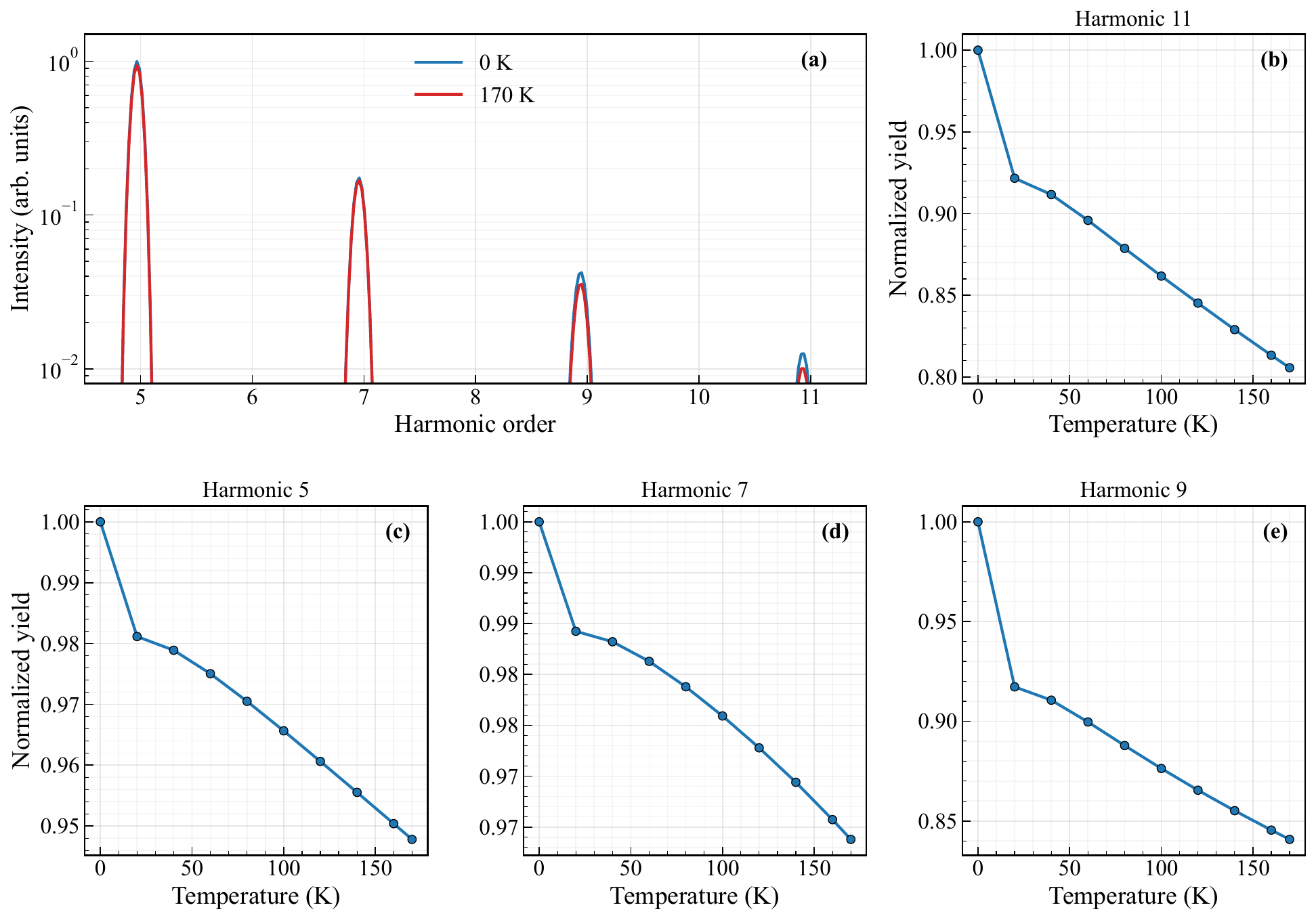}
\caption{
Temperature dependence of the high-harmonic response including quantum lattice fluctuations. 
(a) High-harmonic spectra computed at $0$ K and $170$ K using thermally distorted lattice configurations generated from the quantum phonon displacement amplitudes. The distortions are constructed from the same four optical phonon modes discussed in the main text. 
(b–e) Temperature dependence of the normalized yield for selected harmonics (11$^{th}$, 5$^{th}$, 7$^{th}$, and 9$^{th}$ orders).
}
\label{fig:HHG_quantum}
\end{figure}

Using these temperature-dependent amplitudes, we generate ensembles of distorted lattice configurations and evaluate the corresponding high-harmonic spectra. 
The resulting spectra and harmonic yields are shown in Fig.~S\ref{fig:HHG_quantum}. 
%As the phonon occupation increases with temperature, the associated lattice distortions become larger, leading to a systematic modification of the harmonic response. 
Notably, the reduction in harmonic yield obtained using the quantum phonon distribution is significantly weaker than in the classical case. 
In both calculations we include the same four optical phonon modes; however, in the quantum description the phonon amplitudes are governed by the Bose-Einstein occupation factor, which increases more gradually with temperature than the classical equipartition scaling used in the classical model. 
As a result, the lattice distortions remain smaller and the impact on the harmonic emission is correspondingly reduced.
We emphasize that only four phonon modes are included in the present calculations, whereas a realistic thermal lattice would involve contributions from many more modes. Nevertheless, both approaches consistently show that the harmonic yield decreases with increasing temperature, that the reduction becomes more pronounced for higher harmonics, and that this behavior originates from phonon-induced lattice distortions that disrupt the coherent electronic dynamics responsible for high-harmonic generation.

\renewcommand{\figurename}{Supplementary Fig.}

%In the high-temperature limit $k_B T \gg \hbar\Omega_\nu$, the Bose--Einstein distribution reduces to
%\begin{equation}
%n_B(\Omega_\nu,T) \approx \frac{k_B T}{\hbar\Omega_\nu},
%\end{equation}
%and the quantum expression for the displacement amplitude recovers the classical equipartition result derived above.

\newpage
\section{Incoherent ensemble averaging of high-harmonic spectra and yields}
\begin{figure}[h!]
\centering
\includegraphics[width=0.8\linewidth]{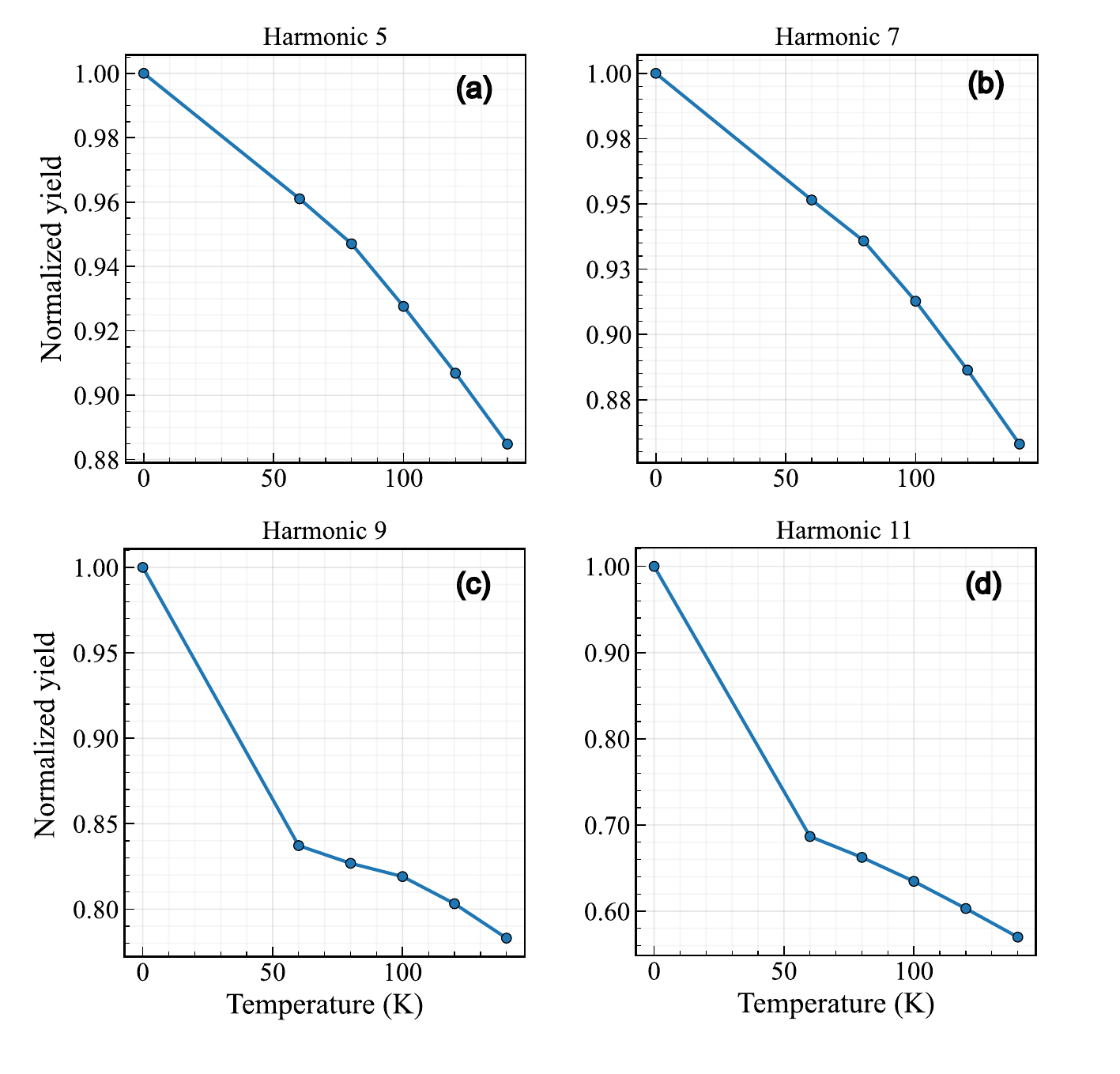}
\caption{
Temperature dependence of normalized high-harmonic yields obtained using an incoherent ensemble average over thermally distorted lattice configurations. Panels (a)-(d) show the yields of harmonics 5, 7, 9, and 11, respectively, normalized to their values at $T=0$~K.
 }
\label{fig:HHG_incoherent}
\end{figure}

To further clarify the role of configuration-dependent phase variations, we also compute the harmonic response using an incoherent ensemble sum over thermally distorted configurations. 
In this case, the harmonic intensities from different configurations are averaged without retaining their relative phases. The resulting temperature dependence of the harmonic yields is shown in Fig.~S\ref{fig:HHG_incoherent}. 
In contrast to the coherent summation presented in the main text, the incoherent sum exhibits a noticeably weaker reduction of the harmonic yield with increasing temperature. 
This behavior indicates that a significant part of the suppression observed in the coherent case originates from configuration-dependent phase variations that lead to partial destructive interference between different lattice realizations. 
When these phase relations are neglected, the remaining reduction primarily reflects the diminished emission strength of individual distorted geometries.

\renewcommand{\figurename}{Supplementary Fig.}

\newpage
\section{Polarization dependence}
\renewcommand{\figurename}{Supplementary Fig.}
\begin{figure}[h!]
\centering
\includegraphics[width=0.8\linewidth]{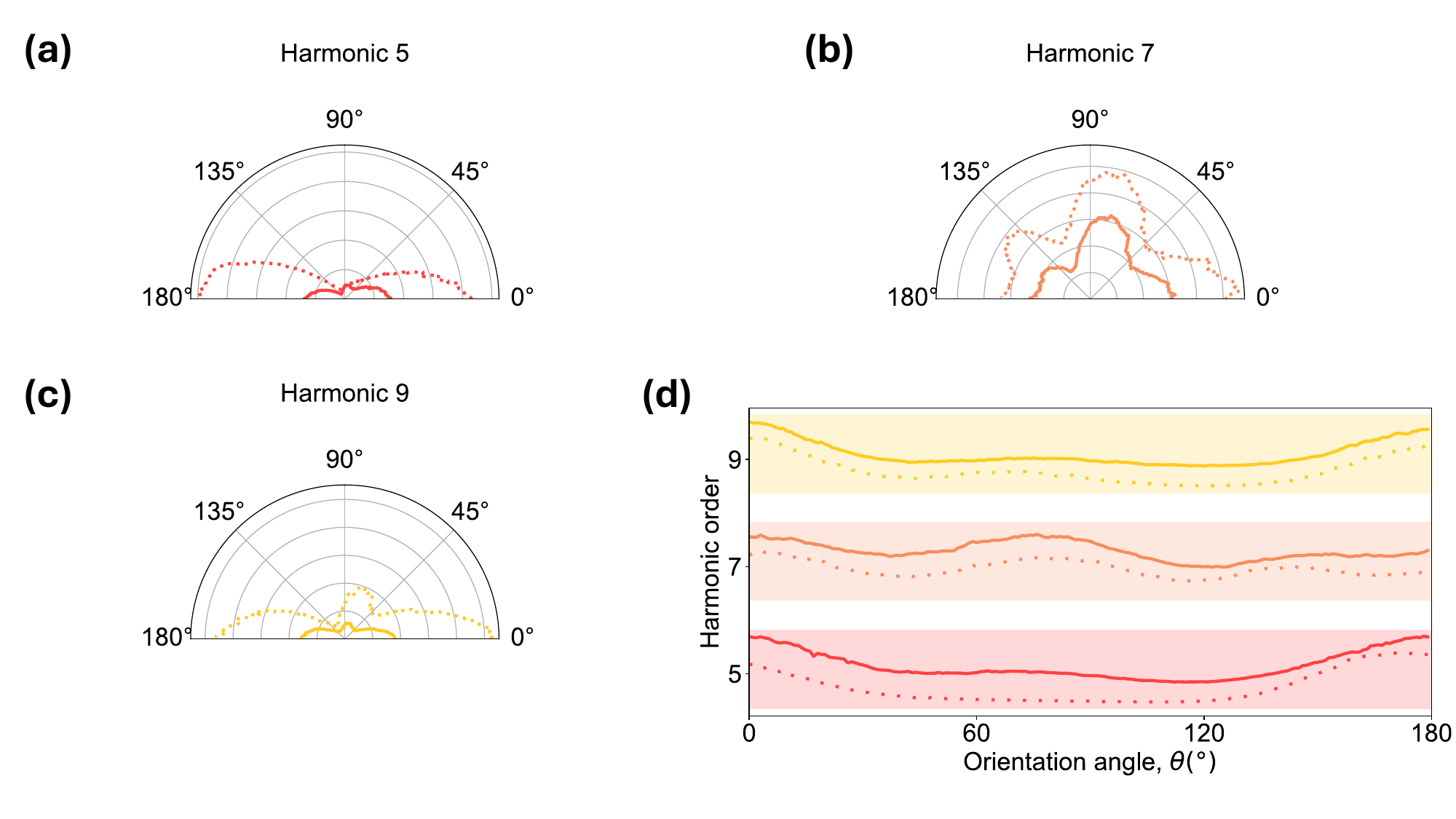}
\caption{Dependence of HHG on laser polarization for Re$_6$Se$_8$Cl$_2$. (a-c) Orientation dependence of 5th, 7th, and 9th harmonics respectively recorded at 290 K (solid line) and 7 K (dotted line) in polar coordinates. (d) Representation of 5th (red), 7th (orange), and 9th (yellow) in cartesian coordinates recorded at 290 K (solid line) and 7 K (dotted line). The magnitude of each on the y-axis does not represent the magnitude of the recorded harmonics at those temperatures. } 
\label{fig:Polarization_dependence}
\end{figure}
  We measured high-harmonic yield as a function of crystal orientation and showed that high-harmonic enhancement at low temperatures applies to all crystal orientations. Experimentally, we keep the driving field linearly polarized and change its direction with respect to the crystal a-axis. We performed these measurements at 290 K and 7 K (Fig.~S\ref{fig:Polarization_dependence}). For all harmonic orders, we observed a pronounced dependence of the harmonic yield on laser polarization, which is likely due to anisotropic band structures involved in high-harmonic generation. Apart from an overall harmonic enhancement at 7 K, the angular dependence remains qualitatively the same at 290 K and 7 K. This observation suggests that there is no drastic electronic state or lattice symmetry change at low temperatures. These results are consistent with reduced thermal lattice fluctuation at all orientations.
  
\newpage
\section{Temperature-dependent reflectivity of the driving field}
\renewcommand{\figurename}{Supplementary Fig.}
\begin{figure}[h!]
\centering
\includegraphics[width=0.8\linewidth]{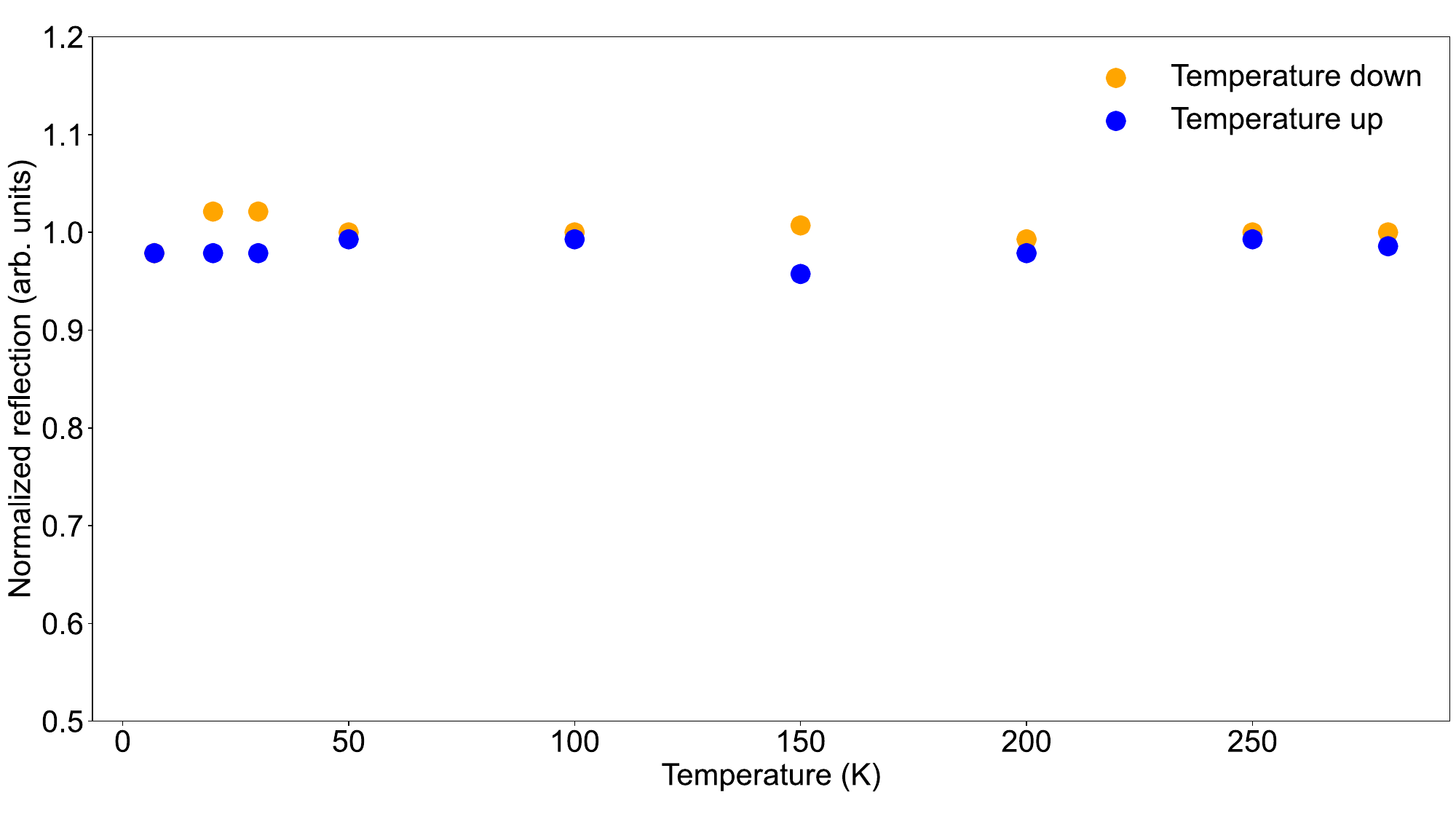}
\caption{Reflectivity of the fundamental driving field (0.35 eV) as a function of sample temperature. Orange circles were taken as the sample temperature was lowered, while the blue circles were taken as the temperature was raised. The reflectivity does not show a clear temperature dependence.} 
\label{fig:fundamental reflection}
\end{figure}
To rule out the contribution from the nonlinear absorption in the observed high-harmonic enhancement, we measured the reflected 3.5 $\mu$m driving pulse energy after high-harmonic generation at different sample temperatures. Experimentally, the reflected 3.5 $\mu$m beam is isolated from harmonic photons and measured by a mid-infrared photodiode. Fig.~S\ref{fig:fundamental reflection} shows the reflected 3.5 $\mu$m pulse energy at different sample temperatures, normalized to the reflectivity at 280 K. The data is collected on both sample cooling and heating cycles. Importantly, the reflectivity largely remains constant and does not exhibit a clear temperature dependence. Based on this measurement, we conclude that the observed high-harmonic enhancement at low temperatures is not due to the nonlinear absorption of the driving field.

\newpage
\bibliography{solid_HHG}